\documentclass[a4paper,11pt]{article}

\usepackage{cite}

\usepackage{authblk}

\usepackage{amsmath}
\usepackage{amscd}
\usepackage{amsfonts}
\usepackage{amssymb}

\begin{document}
\title{\bf{Poincar\'{e} gauge symmetries, hamiltonian symmetries and trivial gauge transformations\vspace{1em}}}
\author{\textsl{Rabin Banerjee}\thanks{\texttt{rabin@bose.res.in}}}
\author{\textsl{Debraj Roy}\thanks{\texttt{debraj@bose.res.in}}}
\affil{\it\small S.~N.~Bose National Centre for Basic Sciences,\\ \it\small Block-JD, Sector III, Salt Lake, Kolkata-700098, India.}

\date{}

\maketitle

\vspace{-1em}
\begin{abstract}
We resolve a problem of finding the Poincar\'{e} symmetries from hamiltonian gauge symmetries constructed through a canonical procedure of handling constrained systems. Through the use of Noether identities corresponding to the symmetries, we motivate a procedure of finding the map between the hamiltonian and Poincar\'{e} gauge parameters. Using this map, we show that the Poincar\'{e} and hamiltonian gauge symmetries are equivalent, modulo trivial gauge transformations.
\end{abstract}

\section{Introduction}
\label{Sec:Intro}

Gauge symmetries in various diffeomorphism invariant theories are a matter of continued interest. Some, among the multitude of models where gauge symmetries have been studied, are Chern-Simons gauge theory \cite{Witten:1988hc}, Einstein-Cartan gravity \cite{Blagojevic:2002du,Frolov:2009wu}, topological gravity with torsion \cite{Blagojevic:2004hj,Banerjee:2009vf} and topologically massive gravities \cite{Deser:1981wh,Blagojevic:2008bn} including Bergshoeff-Holm-Townsend (BHT) or ``new massive gravity'' \cite{Blagojevic:2010ir,Blagojevic:2011qc,Banerjee:2011rx}. By gauge symmetries we mean those transformations of the basic fields of the action, parametrised by arbitrary functions of time, that leave the action invariant under appropriate boundary conditions \cite{Henneaux:1992ig}. The arbitrary functions of time are the gauge parameters.

On the other hand, diffeomorphism invariant theories have the Poincar\'{e} symmetries `$\delta_{\scriptscriptstyle PGT}$', i.e. local Lorentz rotations and translations, as off-shell symmetries by construction. They are found in the usual manner of gauge theories, through localisation of the Poincar\'{e} symmetries \cite{Utiyama:1956sy,Kibble:1961ba,Sciama:1962,Hehl:1976kj,Blagojevic:2002du} and so their form dosen't depend on the particular diffeomorphism invariant model being considered. Say, for example, let us first consider the Einstein-Cartan action in 3D $$ S_1 = \int \mathrm{d^3x} ~\epsilon^{\mu\nu\rho}\, b^i_{\ \mu}R_{i\nu\rho}\,,$$ and then add to it the torsion $T_{i\nu\rho}$ enforced by a parameter $\alpha_4$ $$ S_2 = \int \mathrm{d^3x} ~\epsilon^{\mu\nu\rho} \left[ b^i_{\ \mu}R_{i\nu\rho} + \frac{\alpha_4}{2}\, b^i_{\ \mu}T_{i\nu\rho}\right]$$ where $b^i_{\ \mu}$ is the triad and $R_{i\nu\rho}$ is the Riemann tensor. The Poincar\'{e} symmetry of the (for example) triad field is the same for both of these actions $$ \delta_{\scriptscriptstyle PGT} b^i_{\ \mu} = -\epsilon^i_{\ jk}b^j_{\ \mu}\theta^k - \partial_\mu \xi^\rho \,b^i_{\ \rho} - \xi^\rho\,\partial_\rho b^i_{\ \mu} $$ and as we can see, it does not involve the coupling constant $\alpha_4$. It is off-shell by construction and this can be easily checked explicitly \cite{Banerjee:2009vf}. The gauge parameters here are $\xi^\rho$ for translations and $\theta^i$ for local Lorentz rotations.

To study the hamiltonian gauge symmetries `$\delta_{\scriptscriptstyle G}$', canonical hamiltonian analysis of all the models mentioned above has been carried out extensively, in the literature cited above (also see references therein). The nature of the hamiltonian symmetries depend intimately on the particular model being studied, through the structural nature of the constraints. These hamiltonian symmetries reveal a striking feature, in all of the models. The Poincar\'{e} symmetries are not identifiable from the hamiltonian gauge symmetries. For example, for the Einstein-Cartan action with torsion we get $$\delta_{\scriptscriptstyle G} b^i_{\ \mu} = \nabla_\mu\varepsilon^i + \alpha_4 \,\epsilon^i_{\ jk} \,b^j_{\ \mu} \varepsilon^k + \epsilon^i_{\ jk}\,b^j_{\ \mu} \tau^k\,,$$ where $\varepsilon^i$ and $\tau^i$ are the gauge parameters.  Note that $\delta_{\scriptscriptstyle G}$ explicitly involves the coupling constant $\alpha_4$. To compare $\delta_{\scriptscriptstyle PGT}$ and $\delta_{\scriptscriptstyle G}$ we first have to map the (arbitrary) gauge parameters of the hamiltonian symmetries $\delta_{\scriptscriptstyle G}$ to those of the Poincar\'{e} symmetries $\delta_{\scriptscriptstyle PGT}$. The gauge parameters become different as the Poincar\'{e} parameters are dictated by either geometric or group theoretic demands while the hamiltonian parameters depend on the structure of the constraints arising in the theory. The required redefinition is usually done through an ad-hoc, field-dependant map \cite{Witten:1988hc,Blagojevic:2002du,Blagojevic:2004hj,Banerjee:2009vf,Blagojevic:2008bn,Blagojevic:2010ir,Blagojevic:2011qc,Banerjee:2011rx}. After such a mapping it is seen that the hamiltonian symmetries indeed give back the Poincar\'{e} symmetries, but modulo terms proportional to equations of motion \cite{Blagojevic:2004hj,Banerjee:2009vf}. $$\delta_{\scriptscriptstyle G} b^i_{\ \mu} \sim \delta_{\scriptscriptstyle PGT} b^i_{\ \mu} + \text{\sl Eqns.  of  motion}$$ So the hamiltonian symmetries are not exactly equal to the Poincar\'{e} symmetries and it seems that we may have two independent sets of symmetries for the same action! Each of these symmetries will now give rise to their own independent Noether identities.

This is not a desirable situation. It leads to an increase in the total number of independent gauge  parameters over and above that found through the canonical analysis. We now have to take the Poincar\'{e} symmetry parameters in addition to the hamiltonian gauge parameters, if they are distinct. Also, we have to deal with more number of independent Noether identities than the number of Poncar\'{e} symmetries present in the model. But we know that the number of gauge parameters and Noether identities must match the total number of independent, primary, first-class constraints \cite{Henneaux:1990au,Banerjee:1999yc,Banerjee:1999hu}. This creates an apparent paradoxical situation.

In this article we provide a resolution of this paradox by pointing out that the pair of symmetries differ only through trivial gauge transformations. These types of transformations \cite{Henneaux:1992ig} are not generated by first-class constraints of the theory. Thus they do not introduce any new arbitrary functions of time, i.e. they give rise to no new gauge parameters. Hence they indicate no degeneracy in the equations of motion and their solutions, representing physical states, are not mapped to new degenerate states through these transformations. Thus trivial gauge symmetries are not physical and the hamiltonian mechanism actually reproduce the Poinacr\'{e} symmetries as the only physically relevant symmetries of the theory. Such symmetries also produce no new independent Noether identity and so the total number of identities and gauge parameters match the original number of Poincar\'{e} symmetries. By exploiting the Noether identities we provide a method to construct the map between the hamiltonian and Poincar\'{e} gauge parameters. Finally, through this work, explicit examples of trivial gauge symmetries and the role they play in many well studied field theories get highlighted.

We now explain the organisation of our article. In section \ref{Sec:Trivial} we give a brief overview of trivial gauge symmetries from a general point of view, including a discussion on their role in the Noether identities. In section \ref{Sec:EC}, we take up the first order generalisation of Einstein gravity -- the Einstein-Cartan model. We explicitly show how the two sets of symmetries and Noether identities corresponding to hamiltonian and PGT formulations are related via trivial gauge transformations. We also motivate an algorithm to find suitable map between gauge parameters to enable a comparison between these two sets of symmetries. In section \ref{Sec:MB}, the analysis is performed in a generalisation of the previous model to a Mielke-Baekler type gravity \cite{Mielke:1991nn,Baekler:1992ab} extended by a cosmological term. This provides a further non-trivial demonstration of our results. Finally, we conclude in section \ref{Sec:Disc}.

\noindent{\em Summary of conventions:} Latin indices refer to the local Lorentz frame and the Greek indices refer to the coordinate frame. The beginning letters of both alphabets $(a,b,c,\ldots)$ and $(\alpha,\beta,\gamma,\ldots)$ run over the space part (1,2) while the middle alphabet letters $(i,j,k,\ldots)$ and $(\mu,\nu,\lambda,\ldots)$ run over all coordinates (0,1,2). The totally antisymmetric tensor $\epsilon^{ijk}$ and the tensor density $\epsilon^{\mu\nu\rho}$ are both normalized so that $\epsilon^{012}=1$. The signature of space-time adopted here is $\eta = \text{diag}(+,-,-)$.

\section{Gauge symmetries and trivial gauge symmetries}
\label{Sec:Trivial}

Let $S[q_i]$ describe an action with the basic field variables being $q_i$ $(i=1,2,\ldots,n)$. The canonical momenta are then defined as $\pi^i=\frac{\delta S}{\delta \dot{q_i}}$ and the hamiltonian phase space is constructed out of the conjugate pair $(q_i, \pi^i)$. The standard canonical procedure \cite{Dirac:Lectures} yields all the constraints. Let us denote the first class constraints as $\Sigma_{a}$, $(a=1,2,\ldots,f)$ and the second class constraints as $\chi_{b}$ $(b=1,2,\ldots,s)$, with $P=f+s$ being the total number of constraints. The Dirac prescription gives the gauge generator as a linear combination of all first class constraints $$G= \alpha^a\,\Sigma_{a},$$ $\alpha^a$'s being arbitrary parameters in time. However, not all the parameters $\alpha^a$ are independent. We can eliminate the dependant ones systematically and write the gauge generator in terms of only the independent $\alpha^a$'s, following a completely off-shell method \cite{Henneaux:1990au,Banerjee:1999yc,Banerjee:1999hu}.\footnote{There are other methods of construction of a gauge generator like \cite{Castellani:1981us}, though it is not an off-shell one.} The final generator yields the gauge transformations of fields through a Poisson bracket\footnote{Or a Dirac bracket, if the second class sector has been eliminated through introduction of Dirac brackets.} operation with the fields. There exist two different possibilities of defining this operation $\lbrace q, G\rbrace$, results being equivalent upto terms proportional in constraints.
\begin{align}
\label{transform Poisson}
\begin{aligned}
\delta_1 q &= \lbrace q, \alpha^a\,\Sigma_a \rbrace \\
\text{or,}\quad \delta_2 q &= \alpha^a \lbrace q, \Sigma_a \rbrace.
\end{aligned}
\end{align}
These two definitions of gauge transformations $\delta_1$ and $\delta_2$ differ upto `trivial gauge transformations' \cite{Henneaux:1990au}.

Trivial gauge transformations keep the action invariant simply by a specific antisymmetric structure within them. To write explicitly, let us consider transformations of the form
\begin{align}
\label{trivial gauge gen}
\delta q_i = \Lambda_{ij}\,\frac{\delta S}{\delta q_j},\qquad \Lambda_{ij}=-\Lambda_{ji}.
\end{align}
Here $\frac{\delta S}{\delta q_j}$ is the Euler derivative corresponding to the field $q_j$ and its equation of motion is given by setting this Euler derivative to zero. Thus on-shell, i.e. after imposition of equations of motion, trivial gauge transformations vanish. However invariance of the action $(\delta S = 0)$ is achieved off-shell due to the antisymmetry of $\Lambda_{ij}$
\begin{align}
\label{var action for trivial}
\delta S &= \frac{\delta S}{\delta q_i} \delta q_i \nonumber\\
&= \frac{\delta S}{\delta q_i}\,\Lambda_{ij}\,\frac{\delta S}{\delta q_j} = 0,
\end{align}
as the product $\frac{\delta S}{\delta q_i}\,\frac{\delta S}{\delta q_j}$ is symmetric in $i \,\&\, j$.
Since these transformations vanish on-shell they imply no degeneracy in the solutions of the equations of motion; i.e. they do not map a set of solutions to any other set through arbitrary functions of time, unlike true gauge transformations. Given any action, they can always be added as symmetry transformations and the specific form of the co-efficients do not matter, as long as they are antisymmetric in the field indices. They are not generated by first-class constraints in the hamiltonian formalism and give rise to zero gauge current as they are on-shell symmetries. Thus, trivial gauge symmetries are not true gauge symmetries and are of no physical importance.

As a consequence of the above discussion, it can be anticipated that the trivial gauge symmetries do not give rise to any new Noether identities, other than those already present due to the true symmetries of the system. Given any gauge symmetry parametrised by an arbitrary time function $\sigma$ (known as the gauge parameter), $$\delta q_i = R_{i\mu} \sigma^\mu +  \tilde{R}_{i\mu}^{\ \ \nu}\,\partial_\nu \sigma^\mu$$ where $R_i$'s and $\tilde{R}_i$'s are functions of the fields $q_i$ and possibly their derivatives, the invariance of the action leads to 
\begin{align}
\label{Noether Id gen}
\delta S &= \int \frac{\delta \mathcal{L}}{\delta q_i} \delta q_i \nonumber\\
&= \int \frac{\delta \mathcal{L}}{\delta q_i} \left(R_{i\mu} \sigma^\mu +  \tilde{R}_{i\mu}^{\ \ \nu}\,\partial_\nu \sigma^\mu \right) \nonumber\\
&= \int \left[\frac{\delta \mathcal{L}}{\delta q_i}\,R_{i\mu} - \partial_\nu\left( \frac{\delta \mathcal{L}}{\delta q_i} \, \tilde{R}_{i\mu}^{\ \ \nu} \right) \right]\sigma^\mu = 0.
\end{align}
Since $\sigma$ is an arbitrary function, we can write
\begin{align}
\label{Noether Id gen contd}
\frac{\delta \mathcal{L}}{\delta q_i}\,R_{i\mu} - \partial_\nu\left( \frac{\delta \mathcal{L}}{\delta q_i} \, \tilde{R}_{i\mu}^{\ \ \nu} \right) = 0
\end{align}
which are the Noether identities of the system\footnote{For a lagrangian analysis based on these identities, in the context of the Mielke-Baekler model, see \cite{Banerjee:2010kd}.}. They imply a dependence of the Euler derivatives $\frac{\delta \mathcal{L}}{\delta q_i}$ among themselves and thus the equations of motion are not all independent. Note that each Noether identity is proportional to a gauge parameter (here $\sigma^\mu$). Thus combinations of one set of independent  Noether identities among themselves to give rise to another equivalent set of identities is reflected at the symmetry level as a redefinition of the old gauge parameters into a new set of gauge parameters.

Now trivial gauge symmetries may affect the Noether identities in many ways. In a direct manner, if $R_i$ has antisymmetric contributions like $$R_{i\mu} \rightarrow R_{i\mu} + (\Lambda_{ij})_\mu\,\frac{\delta \mathcal{L}}{\delta q_j}\,\qquad(\Lambda_{ij})_\mu=-(\Lambda_{ji})_\mu,$$ as can arise from transformations like \eqref{trivial gauge gen}, then we will have extensions of the gauge identities \eqref{Noether Id gen contd} as shown below
\begin{align}
\label{Trivial Noether Id gen}
\frac{\delta \mathcal{L}}{\delta q_i}\,R_{i\mu} - \partial_\nu\left( \frac{\delta \mathcal{L}}{\delta q_i} \, \tilde{R}_{i\mu}^{\ \ \nu} \right) + \frac{\delta \mathcal{L}}{\delta q_i} (\Lambda_{ij})_\mu\,\frac{\delta \mathcal{L}}{\delta q_j} = 0.
\end{align}
However the last term vanishes by itself, without depending on the particular structure of the Euler derivatives, through (anti)symmetry. This generates no new identities and thus the Noether identities \eqref{Noether Id gen contd} and \eqref{Trivial Noether Id gen} are infact equivalent to each other and correspond to only one {\em physical} symmetry.

In the following sections, we work with explicit models (Einstein-Cartan gravity and a Mielke-Baekler \cite{Mielke:1991nn,Baekler:1992ab} type gravity) to show the role of trivial gauge symmetries in relating hamiltonian symmetries to the Poincar\'{e} symmetries. The analysis in each case will be based on the general formalism outlined in this section.

\section{Einstein -- Cartan gravity}
\label{Sec:EC}

The Einstein-Cartan formulation of gravity is a first order generalisation of Einstein's general relativity. It is constructed through a Poincar\'{e} gauge theory (PGT) construction,  \cite{Utiyama:1956sy,Kibble:1961ba,Sciama:1962,Hehl:1976kj,Blagojevic:2002du} on a Riemann-Cartan spacetime having both curvature, as well as torsion. To start with, triad fields $b^i_{\ \mu}(x)$ are set up at each point of spacetime to translate between local coordinates $x^i$ and global coordinates $x^\mu$. Thus, for any vector $A_{\mu}$, we have $A_\mu = b^i_{\ \mu} A_i$. The global metric $g_{\mu\nu}$ is written in terms of the triads and the local flat Minkowski metric $\eta^{ij}$ as $$g_{\mu\nu} = b^i_{\ \mu} b^j_{\ \nu}\, \eta_{ij}\,.$$ At this stage, there is a global Poincar\'{e} symmetry parametrised by Lorentz rotations and translations. To localise this Poincar\'{e} symmetry, covariant derivatives are brought in to replace partial derivatives and spin-connection fields $\omega^i_{\ \mu}$ are introduced. These comprise independent fields in PGT. The corresponding field strengths that come into play through the commutators of covariant derivatives give rise to torsion $T^i_{\ \mu\nu}$ and curvature $R^i_{\ \mu\nu}$ tensors. Their forms turn out to be
\begin{align}
\label{PGT R T}
\begin{aligned}
R^i_{\ \mu\nu} &= \partial_\mu \omega^i_{\ \nu} - \partial_\mu \omega^i_{\ \nu} + \epsilon^i_{\ jk}\,\omega^j_{\ \mu}\omega^k_{\ \nu} \\
T^i_{\ \mu\nu} &= \nabla_\mu b^i_{\ \nu} - \nabla_\nu b^i_{\ \mu}.
\end{aligned}
\end{align}
Here the covariant derivative of the triad is defined as $\nabla_\mu  b^i_{\ \nu} = \partial_\mu b^i_{\ \nu} + \epsilon^i_{\ jk}\,\omega^j_{\ \mu}b^k_{\ \nu}$. The PGT gravity models are constructed to be invariant under the local Poincar\'{e} transformations
\begin{align}
\label{PGT deltas}
\begin{aligned}
\delta_{\scriptscriptstyle PGT} b^i_{\ \mu} &= -\epsilon^i_{\ jk}b^j_{\ \mu}\theta^k - \partial_\mu \xi^\rho \,b^i_{\ \rho} - \xi^\rho\,\partial_\rho b^i_{\ \mu} \\
\delta_{\scriptscriptstyle PGT} \omega^i_{\ \mu} &= -\partial_\mu \theta^i - \epsilon^i_{\ jk}\omega^j_{\ \mu}\theta^k - \partial_\mu\xi^\rho\,\omega^i_{\ \rho} - \xi^\rho\,\partial_\rho\omega^i_{\ \mu}.
\end{aligned}
\end{align}
In the above symmetries, the parameter describing local Lorentz transformations is $\theta^i$ and that describing general coordinate transformations is $\xi^\mu$ (both transformations being of infinitesimal order). Intuitively, this explains the structure of the transformations \eqref{PGT deltas} where the index `$i$' transforms as a Lorentz index while `$\rho$' transforms as a general coordinate index.\footnote{For a more detailed discussion one may refer to \cite{Blagojevic:2004hj,Banerjee:2009vf}.} The number of independent Poincar\'{e} symmetries for each field $(b,\ \omega, \text{ or any other field, if present})$ is reflected in the number of independent gauge parameters. In our model in 3D, $i=0,1,2$ and $\rho=0,1,2$. So the total number is
\begin{align}
\label{counting PGT}
3 \text{ against } \xi^\rho + 3 \text{ against } \theta^i = 6.
\end{align}
So we expect to find $6$ independent gauge parameters and $6$ independent Noether identities in our model and no more.

The Einstein-Cartan theory in Riemann-Cartan spacetime gives back the standard Einstein gravity on imposition of the zero torsion condition. The action, in 3D, is
\begin{align}
\label{action EC}
S = \int \textrm{d$^3$x} ~a\,\epsilon^{\mu\nu\rho}\,b^i_{\ \mu}\,R_{i\nu\rho}.
\end{align}
The basic variables of the theory are $b^i_{\ \mu} \text{ and } ~\omega^i_{\ \mu}$ with the corresponding conjugate momenta being denoted by  $\pi_i^{\ \mu} \text{ and } ~\Pi_i^{\ \mu}$ respectively. The variational equations of motion are given by setting the Euler derivatives $\frac{\delta S}{\delta b^i_{\ \mu}}$ and $\frac{\delta S}{\delta \omega^i_{\ \mu}}$ to zero.
\begin{align}
\label{EOM EC}
\begin{aligned}
\frac{\delta S}{\delta b^i_{\ \mu}} &= a\,\epsilon^{\mu\nu\rho}\,R_{i\nu\rho} = 0\\
\frac{\delta S}{\delta \omega^i_{\ \mu}} &= a\,\epsilon^{\mu\nu\rho}\,T_{i\nu\rho} = 0
\end{aligned}
\end{align}
\begin{table}[h]
\centering
\begin{tabular}{l | c c}
\hline\hline\
& First Class & Second class \\[0.2ex] \hline\\[-1.9ex]
Primary & $\phi_i^{\ 0}\;, \Phi_i^{\ 0}$ & $\phi_i^{\ \alpha}$, $\Phi_i^{\ \alpha}$ \\[0.4ex]
Secondary & $\bar{\mathcal{H}}_i\;, \bar{\mathcal{K}}_i$ & \\[0.4ex]
\hline\hline
\end{tabular}
\caption{Constraints of the EC theory.} \label{Tab:Constraints EC}
\end{table}
A Dirac canonical analysis leads to the constraint structure \cite{Blagojevic:2004hj,Banerjee:2009vf} as given in Table \ref{Tab:Constraints EC}.
The relevant quantities in Table \ref{Tab:Constraints EC} are defined below:
\begin{align}
\label{Rel qtys EC}
\begin{aligned}
\phi_i^{\ \mu} &= \pi_i^{\ \mu}\\
\Phi_i^{\ \mu} &= \Pi_i^{\ \mu} - 2a\,\epsilon^{0\alpha\beta}\,b_{i\beta}\delta^\mu_\alpha\\
\bar{\mathcal{H}}_i &= \left[ -a\,\epsilon^{0\alpha\beta}R_{i\alpha\beta} \right] - \nabla_{\!\alpha}\phi_i^{\ \alpha}\\
\bar{\mathcal{K}}_i &= \left[ -a\,\epsilon^{0\alpha\beta}T_{i\alpha\beta} \right] - \nabla_{\!\alpha}\Phi_i^{\ \alpha} - \epsilon_{ijk}\,b^j_{\ \alpha}\phi^{k\alpha}\\
\end{aligned}
\end{align}
Once we have the constraints, we can construct the generator through an explicitly off-shell method \cite{Banerjee:1999yc,Banerjee:1999hu}. For Einstein-Cartan gravity, it turns out to be \cite{Banerjee:2009vf}
\begin{align}
\label{Gen EC}
G = {\ } &\dot{\varepsilon}^i\, \pi_i^{\ 0} + \varepsilon^i\left[ \bar{\mathcal{H}}_i - \epsilon_{ijk}\,\omega^j_{\ 0}\pi^{k0} \right] \nonumber\\
{\ } +\,& \dot{\tau}^i\, \Pi_i^{\ 0} + \tau^i\left[ \bar{\mathcal{K}}_i - \epsilon_{ijk} \left( b^j_{\ 0}\pi^{k0} + \omega^j_{\ 0} \Pi^{k0} \right) \right].
\end{align}
The hamiltonian gauge symmetries are calculated from the generator $G$, adopting the second among the definitions in \eqref{transform Poisson}
\begin{align}
\label{symm G EC}
\begin{aligned}
\delta_{\scriptscriptstyle G} b^i_{\ \mu} &= \nabla_\mu \varepsilon^i + \epsilon^i_{\ jk}b^j_{\ \mu} \tau^k\\
\delta_{\scriptscriptstyle G} \omega^i_{\ \mu} &= \nabla_\mu \tau^i .
\end{aligned}
\end{align}
Now the generator \eqref{Gen EC} is constructed as a linear combination of the products of first class constraints with gauge parameters. Looking at the first-class constraints in Table \ref{Tab:Constraints EC}, we see that they all have one local index as their free-index. This fixes the structure of the gauge parameters $\varepsilon^i$ and $\tau^i$ in the hamiltonian formulation and they turn out to be different from the Poincar\'{e} gauge parameters $\xi^\rho$ and $\theta^i$, translations and local Lorentz rotations, seen in \eqref{PGT deltas}. However, to compare between the two symmetries $\delta_{\scriptscriptstyle G}$ and $\delta_{\scriptscriptstyle PGT}$ we must first have structurally similar set of gauge parameters in both sets of symmetries. This is achieved by introducing a field dependant map between the hamiltonian and Poincar\'{e} gauge parameters \cite{Blagojevic:2002du,Blagojevic:2004hj,Banerjee:2009vf}
\begin{align}
\label{map1}
\varepsilon^i = -\xi^\rho\,b^i_{\ \rho} \qquad\&\qquad \tau^i = -\theta^i - \xi^\rho \omega^i_{\ \rho}.
\end{align}
But this map is usually proposed arbitrarily and there is no process to generate this map from physical considerations. Using this map in the symmetries \eqref{symm G EC} and after a bit of manipulations, we arrive at
\begin{align}
\label{symm mapped EC}
\begin{aligned}
\delta_{\scriptscriptstyle G} b^i_{\ \mu} &= \delta_{\scriptscriptstyle PGT}b^i_{\ \mu} + \frac{1}{2a}\,\xi^\rho\,\epsilon_{\mu\nu\rho}\,\frac{\delta S}{\delta \omega_{i\nu}} \\
\delta_{\scriptscriptstyle G} \omega^i_{\ \mu} &= \delta_{\scriptscriptstyle PGT}\omega^i_{\ \mu} + \frac{1}{2a}\,\xi^\rho\,\epsilon_{\mu\nu\rho}\,\frac{\delta S}{\delta b_{i\nu}},
\end{aligned}
\end{align}
where the Euler derivatives are defined in \eqref{EOM EC}. So the two sets of symmetries are different, and match only on-shell. Consequently, they also give rise to two sets of Noether identities.

The Noether identities corresponding to the PGT symmetries \eqref{PGT deltas} can be found by proceeding along the route leading to \eqref{Noether Id gen contd}. Explicitly, they are \cite{Banerjee:2009vf}
\begin{subequations}
\label{Noether PGT EC}
\begin{align}
\label{Noether PGT EC 1}
P_k &= \frac{\delta S}{\delta b^i_{\ \mu}} \varepsilon^i_{\ jk} b^j_{\ \mu} + \frac{\delta S}{\delta \omega^i_{\ \mu}} \varepsilon^i_{\ jk} \omega^j_{\ \mu} -\partial_\mu\left(\frac{\delta S}{\delta \omega^k_{\ \mu}}\right) = 0\\
\label{Noether PGT EC 2}
R_\rho &= \frac{\delta S}{\delta b^i_{\ \mu}} \partial_\rho b^i_{\ \mu} + \frac{\delta S}{\delta \omega^i_{\ \mu}} \partial_\rho \omega^i_{\ \mu} - \partial_\mu \left(b^i_{\ \rho}\frac{\delta S}{\delta b^i_{\ \mu}} + \omega^i_{\ \rho} \frac{\delta S}{\delta \omega^i_{\ \mu}} \right) = 0.
\end{align}
\end{subequations}
The total number is $3+3=6$, as expected. Those corresponding to the hamiltonian gauge transformations \eqref{symm G EC} are, similarily,
\begin{subequations}
\label{Noether gauge EC}
\begin{align}
\label{Noether gauge EC 1}
A_k &= -\partial_\mu\left(\frac{\delta S}{\delta \omega^k_{\ \mu}}\right) + \frac{\delta S}{\delta b^i_{\ \mu}} \varepsilon^i_{\ jk} b^j_{\ \mu} + \frac{\delta S}{\delta \omega^i_{\ \mu}} \varepsilon^i_{\ jk} \omega^j_{\ \mu} = 0\\
\label{Noether gauge EC 2}
B_k &= -\partial_\mu\left(\frac{\delta S}{\delta b^k_{\ \mu}}\right) + \frac{\delta S}{\delta b^i_{\ \mu}} \varepsilon^i_{\ jk} \omega^j_{\ \mu} = 0
\end{align}
\end{subequations}
and are also $3+3=6$ in number. We would like to emphasise at this point that these identities are to be dealt with {\em off-shell}, without imposition of equations of motion, i.e. without setting the Euler derivatives to be zero. The identities in-fact become tautological $0=0$ statements on-shell as they are comprised of the Euler derivatives.

Now the question that we want to address is whether the sets of identities \eqref{Noether PGT EC} and \eqref{Noether gauge EC} are independent, or can they be shown to be actually the same. A comparison shows that among the two sets, \eqref{Noether PGT EC 1} and \eqref{Noether gauge EC 1} are already identical, i.e. $P_k \equiv A_k$. We want to check the possibility of expressing $R_\rho$ as some linear combination of $P_k$ and $R_k$. Comparing the structure of the free indices and the derivative terms among \eqref{Noether PGT EC 1} and \eqref{Noether gauge EC 1} we see that the combination $-b^k_{\ \rho}B_k - \omega^k_{\ \rho}A_k$ gives us
\begin{align}
\label{PGT Gauge Inv EC}
-b^k_{\ \rho}B_k - \omega^k_{\ \rho}A_k = -R_\rho + \frac{\delta S}{\delta b^i_{\ \mu}} \left(\frac{1}{2a}\,\eta^{ij}\,\epsilon_{\mu\nu\rho}\right) \frac{\delta S}{\delta \omega^j_{\ \nu}} +  \frac{\delta S}{\delta \omega^i_{\ \mu}}\left(\frac{1}{2a}\,\eta^{ij}\,\epsilon_{\mu\nu\rho}\right) \frac{\delta S}{\delta b^j_{\ \nu}}=0
\end{align}
The last two terms in the above equation, proportional to square of Euler derivatives, cancel out due to antisymmetry of their coefficients {\em without requiring} the particular form of the Euler derivatives \eqref{EOM EC}. The net identity obtained in the process is just the second Noether identity corresponding to the Poincar\'{e} symmetries. Thus we show that there exists only one set of true, independent Noether identities in the system and the total number of these are $3+3=6$, i.e. equal to the total number of gauge symmetries in the system.

The Noether identities are obtained, as shown in \eqref{Noether Id gen} and \eqref{Noether Id gen contd}, from collecting coefficients of the independent gauge parameters from a variation of the action through functional Taylor expansion
\begin{align}
\label{mapping and Noether EC 1}
\delta S = \int \left( \theta^k P_k + \xi^\rho R_\rho \right) &= 0 \qquad & \text{Poincar\'{e} symmetries.}\\
\label{mapping and Noether EC 2}
\delta S = \int \left( \varepsilon^k A_k + \tau^k B_k \right) &= 0 \qquad & \text{hamiltonian symmetries.}
\end{align}
The combinations $R_\rho = -b^k_{\ \rho}B_k - \omega^k_{\ \rho}A_k$ and $P_k = -A_k$, when substituted in \eqref{mapping and Noether EC 1}, gives $$\int \left[ (-\theta^k - \xi^\rho\omega^k_{\ \rho})\, A_k\, + \,(-b^k_{\ \rho}\xi^\rho)\, B_k \right] = 0.$$ Comparing this with \eqref{mapping and Noether EC 2} gives us the required map \eqref{map1} between the two sets of gauge parameters. So the Noether identities help us to generate the required map between different sets of gauge parameters.

It is desirable to point out that, in the above analysis, we have not used any connection between the Noether identities and equations of motion. A literal application of the dependence of Euler-Lagrange equations due to Noether identities, mentioned below \eqref{Noether Id gen contd}, may lead to incorrect results.\footnote{This point was brought to our notice by the referee who also suggested, in this context, the original classic works of Hilbert on general relativity.} Here we have compared the Noether identities arising from the PGT and hamiltonian approaches to motivate the map \eqref{map1}. Also, all the Noether identities were explicitly verified.

The structure of the antisymmetric terms obtained in \eqref{PGT Gauge Inv EC}, when compared with those that arise in the case of trivial gauge symmetries as outlined in \eqref{Trivial Noether Id gen}, hints at the presence of trivial gauge symmetries within the hamiltonian formalism. The general form of trivial gauge transformations in this model would read
\begin{align}
\label{trivial gauge EC}
\begin{aligned}
\delta b^i_{\ \mu} &= \Lambda_{\left( b^i_{\ \mu},\, b^j_{\ \nu} \right)} \,\frac{\delta S}{\delta b^j_{\ \nu}} + \Lambda_{\left( b^i_{\ \mu},\, \omega^j_{\ \nu} \right)} \,\frac{\delta S}{\delta \omega^j_{\ \nu}} \\
\delta \omega^i_{\ \mu} &= \Lambda_{\left( \omega^i_{\ \mu},\, b^j_{\ \nu} \right)} \,\frac{\delta S}{\delta b^j_{\ \nu}} + \Lambda_{\left( \omega^i_{\ \mu},\, \omega^j_{\ \nu} \right)} \,\frac{\delta S}{\delta \omega^j_{\ \nu}},
\end{aligned}
\end{align}
where $\Lambda$ is antisymmetric (see \eqref{trivial gauge gen}). Here $\delta \equiv \delta_{\scriptscriptstyle G} - \delta_{\scriptscriptstyle PGT}$ is the apparently extra symmetry present within the hamiltonian symmetries. Comparing this with \eqref{symm mapped EC} we find the $\Lambda$ matrix defining the trivial gauge symmetry to be
\begin{align}
\label{Lambda EC}
\begin{aligned}
\Lambda_{\left( b^i_{\ \mu},\, b^j_{\ \nu} \right)} &= 0 \qquad &
	\Lambda_{\left( b^i_{\ \mu},\,\omega^j_{\ \nu} \right)} &= \frac{1}{2a}\,\eta^{ij}\,\xi^\rho\,\epsilon_{\mu\nu\rho} \\
\Lambda_{\left( \omega^i_{\ \mu},\, b^j_{\ \nu} \right)} &= \frac{1}{2a}\,\eta^{ij}\,\xi^\rho\,\epsilon_{\mu\nu\rho} \qquad &
	\Lambda_{\left( \omega^i_{\ \mu},\, \omega^j_{\ \nu} \right)} &= 0
\end{aligned}
\end{align}
The antisymmetry of $\Lambda$ in the diagonal ($b-b$ or $\omega-\omega$) entries is obvious. For the off-diagonal entry,
\begin{align}
\label{antisymm check EC}
\Lambda_{\left( b^i_{\ \mu},\,\omega^j_{\ \nu} \right)} &= \ \frac{1}{2a}\,\eta^{ij}\,\xi^\rho\,\epsilon_{\mu\nu\rho} \nonumber\\
&= - \frac{1}{2a}\,\eta^{ji}\,\xi^\rho\,\epsilon_{\nu\mu\rho} \nonumber\\
&= - \Lambda_{\left( \omega^j_{\ \nu},\, b^i_{\ \mu} \right)}.
\end{align}
Thus the $\Lambda$ matrix is antisymmetric in its field indices and this renders the action off-shell invariant. We have thus shown that the difference between the hamiltonian and Poincar\'{e} symmetries is just a trivial gauge  transformation. The total number of true physical symmetries remain $3+3=6$ as both $\delta_{\scriptscriptstyle G}$ and $\delta_{\scriptscriptstyle PGT}$ are now physically equivalent.

\section{3D cosmological gravity with torsion}
\label{Sec:MB}

In this section, we study a 3D gravity model based on the Mielke-Baekler (MB) action \cite{Mielke:1991nn,Baekler:1992ab} added with a cosmological term. This is formulated with triad -- spin-connection variables, in the PGT formalism. The canonical analysis of this model done in \cite{Blagojevic:2004hj} shows the same feature of hamiltonian and Poincar\'{e} gauge symmetries being related, modulo on-shell vanishing terms.

The action describing this topological 3D gravity model with torsion and a cosmological term is
\begin{align}
\label{action TMG}
S = \int \textrm{d$^3$x}\,\epsilon^{\mu\nu\rho}\!\left[ab^i_{\ \mu}R_{i\nu\rho} - \frac{\Lambda}{3} \epsilon_{ijk}b^i_{\ \mu}b^j_{\ \nu}b^k_{\ \rho} + \alpha_3\!\left(\!\omega^i_{\ \mu}\partial_\nu\omega_{i\rho}  + \frac{1}{3} \epsilon_{ijk}\,\omega^i_{\ \mu}\omega^j_{\ \nu}\omega^k_{\ \rho} \right) + \frac{\alpha_4}{2}b^i_{\ \mu}T_{i\nu\rho} \right]
\end{align}
Basic variables here are $b^i_{\ \mu} \text{ and } \omega^i_{\ \mu}$ and corresponding momenta are denoted $\pi_i^{\ \mu} \text{ and } \Pi_i^{\ \mu}$ respectively. The equations of motion are obtained by setting to zero the various Euler derivatives,
\begin{align}
\label{EOM MB}
\begin{aligned}
\frac{\delta S}{\delta b^i_{\ \mu}} &= \epsilon^{\mu\nu\rho} \left[ a\,R_{i\nu\rho} + \alpha_4\, T_{i\nu\rho} - \Lambda\, \epsilon_{ijk}b^j_{\ \nu}b^k_{\ \rho} \right] = 0 \\
\frac{\delta S}{\delta \omega^i_{\ \mu}} &= \epsilon^{\mu\nu\rho} \left[ \alpha_3\, R_{i\nu\rho} + a\, T_{i\nu\rho} + \alpha_4\, \epsilon_{ijk}b^j_{\ \nu}b^k_{\ \rho} \right] = 0
\end{aligned}
\end{align}
All the momenta turn out to be primary constraints in this first order theory. The consistency process ends at the secondary level itself and the constraints can be classified \cite{Blagojevic:2008bn} as given in Table \eqref{Tab:Constraints MB}.
\begin{table}[h]
\centering
\begin{tabular}{l | c c}
\hline\hline\
& First Class & Second class \\[0.2ex] \hline\\[-1.9ex]
Primary & $\phi_i^{\ 0}\;, \Phi_i^{\ 0}$ & $\phi_i^{\ \alpha}$, $\Phi_i^{\ \alpha} $ \\[0.4ex]
Secondary & $\bar{\mathcal{H}}_i\,, \bar{\mathcal{K}}_i$ &  \\[0.4ex]
\hline\hline
\end{tabular}
\caption{Constraints of the MB type 3D gravity theory.} \label{Tab:Constraints MB}
\end{table}
The relevant quantities used are defined below:
\begin{align}
\label{Rel qtys MB}
\begin{aligned}
\phi_i^{\ \mu} &= \pi_i^{\ \mu} - \alpha_4\,\epsilon^{0\alpha\beta}\, b_{i\beta}\,\delta^\mu_\alpha \\
\Phi_i^{\ \mu} &= \Pi_i^{\ \mu} - \epsilon^{0\alpha\beta} \left( 2 a\, b_{i\beta} + \alpha_3\, \omega_{i\beta} \right) \delta^\mu_\alpha\\
\bar{\mathcal{H}}_i &= - \left[ \epsilon^{0\alpha\beta}\!\left( a\, R_{i\alpha\beta} + \alpha_4\, T_{i\alpha\beta} - \Lambda \epsilon_{ijk} b^j_{\ \alpha} b^k_{\ \beta} \right)\right] - \nabla_{\!\alpha} \phi_i^{\ \alpha} + \epsilon_{ijk}\, b^j_{\ \alpha} \left( p\,\phi^{k\alpha} + q\, \Phi^{k\alpha} \right) \\
\bar{\mathcal{K}}_i &= -\left[ \epsilon^{0\alpha\beta} \left( a\,T_{i\alpha\beta} + \alpha_3\,R_{i\alpha\beta} + \alpha_4\,\epsilon_{ijk} b^j_{\ \alpha} b^k_{\ \beta} \right) \right] - \nabla_{\!\alpha} \Phi_i^{\ \alpha} - \epsilon_{ijk}\, b^j_{\ \alpha} \phi^{k\alpha}\\
p &= \frac{\alpha_3\Lambda + \alpha_4a}{\alpha_3 \alpha_4 - a^2}\,;\qquad q = -\frac{\alpha_4^2 + a\Lambda}{\alpha_3\alpha_4 - a^2}
\end{aligned}
\end{align}
Here the terms within square brackets in the definitions of the constraints $\bar{\mathcal{H}}_i$ and $\bar{\mathcal{K}}_i$, are themselves secondary in nature. The classified constraints in Table \eqref{Tab:Constraints MB} are suitable combinations of the primary and secondary constraints.

Using these constraints and an explicitly off-shell method \cite{Banerjee:1999yc,Banerjee:1999hu}, the hamiltonian generator of gauge symmetries can be constructed \cite{Banerjee:2009vf}. There are two (indexed) gauge parameters $\varepsilon^i$ and $\tau^i$ and they are (again) different from the Poincar\'{e} gauge parameters $\xi^\rho$ and $\theta^k$. The generator `$G$' can be written as a sum of two parts -- $\mathcal{G}_\varepsilon$ and $\mathcal{G}_\tau$, as shown below
\begin{align}
\label{Gen MB}
\begin{aligned}
G=&\int d^2x \left[\mathcal{G}_\varepsilon(x)+\mathcal{G}_\tau(x)\right]\\
&\mathcal{G}_\varepsilon = \dot{\varepsilon}^i\,\pi_i^{\ 0} + \varepsilon^i\left[\bar{\mathcal{H}_i}- \varepsilon_{ijk} \big( \omega^j_{\ 0} - p\,b^j_{\ 0}\big)\pi^{k0} + q \,\varepsilon_{ijk}\,b^j_{\ 0}\Pi^{k0} \right]\\
&\mathcal{G}_\tau = \dot{\tau}^i\Pi_i^{\ 0} + \tau^i\left[\bar{\mathcal{K}_i}-\varepsilon_{ijk}\big(b^j_{\ 0}\,\pi^{k0} + \omega^j_{\ 0}\,\Pi^{k0}\big)\right]
\end{aligned}
\end{align}
Symmetries of the basic fields can be computed from this generator through the second definition among \eqref{transform Poisson}
\begin{align}
\label{symm MB}
\begin{aligned}
\delta b^i_{\ \mu} &= \nabla_\mu\varepsilon^i - p \,\epsilon^i_{\ jk} \,b^j_{\ \mu} \varepsilon^k + \epsilon^i_{\ jk}\,b^j_{\ \mu} \tau^k ,\\
\delta \omega^i_{\ \mu} &= \nabla_\mu \tau^i - q \,\epsilon^i_{\ jk} \,b^j_{\ \mu} \varepsilon^k.\\
\end{aligned}
\end{align}
The hamiltonian symmetries contain the coupling constants $\Lambda, \,\alpha_3 \text{ and } \alpha_4$ through the parameters $p\, \&\, q$ defined earlier. These, they inherit from the action through the structure of the constraints. To compare with Poincar\'{e} symmetries, we take recourse to the map \eqref{map1} relating the hamiltonian gauge parameters to the Poincar\'{e} gauge parameters. After some rearrangements and remembering the Euler derivatives from \eqref{EOM MB}, we arrive at
\begin{align}
\label{symm mapped MB}
\begin{aligned}
\delta_{\scriptscriptstyle G} b^i_{\ \mu} &= \delta_{\scriptscriptstyle PGT} b^i_{\ \mu} + \frac{\alpha_3}{2(\alpha_3\alpha_4-a^2)}\,\eta^{ij}\,\xi^\rho\,\epsilon_{\mu\nu\rho} \,\frac{\delta S}{\delta b^j_{\ \nu}} - \frac{a}{2(\alpha_3\alpha_4-a^2)}\,\eta^{ij}\,\xi^\rho\,\epsilon_{\mu\nu\rho} \,\frac{\delta S}{\delta \omega^j_{\ \nu}} \\
\delta_{\scriptscriptstyle G} \omega^i_{\ \mu} &= \delta_{\scriptscriptstyle PGT} \omega^i_{\ \mu} - \frac{a}{2(\alpha_3\alpha_4-a^2)}\,\eta^{ij}\,\xi^\rho\,\epsilon_{\mu\nu\rho} \,\frac{\delta S}{\delta b^j_{\ \nu}} + \frac{\alpha_4}{2(\alpha_3\alpha_4-a^2)}\,\eta^{ij}\,\xi^\rho\,\epsilon_{\mu\nu\rho} \,\frac{\delta S}{\delta \omega^j_{\ \nu}} \\
\end{aligned}
\end{align}
It is again clear that the hamiltonian and Poincar\'{e} symmetries become identical only on-shell.

Let us now investigate the Noether identities in this model. The identities corresponding to the PGT symmetries remain the same as \eqref{Noether PGT EC}, since the form of the Poincar\'{e} symmetries do not depend upon the form of the lagrangian, as long as the lagrangian is diffeomorphism invariant in nature (and contains the same fields in construction of the action). The hamiltonian gauge symmetries \eqref{symm MB} give rise to the following identities \cite{Banerjee:2009vf}
\begin{subequations}
\label{Noether gauge MB}
\begin{align}
\label{Noether gauge MB 1}
A'_k &= -\partial_\mu\left(\frac{\delta S}{\delta \omega^k_{\ \mu}}\right) + \frac{\delta S}{\delta b^i_{\ \mu}} \varepsilon^i_{\ jk} b^j_{\ \mu} + \frac{\delta S}{\delta \omega^i_{\ \mu}} \varepsilon^i_{\ jk} \omega^j_{\ \mu} = 0\\
\label{Noether gauge MB 2}
B'_k &= -\partial_\mu\left(\frac{\delta S}{\delta b^k_{\ \mu}}\right) + \frac{\delta S}{\delta b^i_{\ \mu}} \varepsilon^i_{\ jk} \omega^j_{\ \mu} -p\, \frac{\delta S}{\delta b^i_{\ \mu}} \epsilon^i_{\ jk} b^j_{\ \mu} - q\, \frac{\delta S}{\delta \omega^i_{\ \mu}} \epsilon^i_{\ jk} b^j_{\ \mu} = 0.
\end{align}
\end{subequations}
Once again we see that one of the identities among the hamiltonian gauge \eqref{Noether gauge MB} and Poincar\'{e} ones \eqref{Noether PGT EC}, $A_k$ and $P_k$, match each other. And the combination $-\omega^k_{\ \rho} A'_k + -b^k_{\ \rho} B'_k$ leads to
\begin{align}
\label{PGT Gauge Inv MB}
-R_\rho + \frac{\delta S}{\delta b^i_{\ \mu}} \left(\frac{\alpha_3}{2(\alpha_3\alpha_4 - a^2)}\,\eta^{ij} \epsilon_{\mu\nu\rho}\right) \frac{\delta S}{\delta b^j_{\ \nu}} + \frac{\delta S}{\delta b^i_{\ \mu}} \left(\frac{-a}{2(\alpha_3\alpha_4 - a^2)}\,\eta^{ij}\, \epsilon_{\mu\nu\rho}\right) \frac{\delta S}{\delta \omega^j_{\ \nu}} & \nonumber\\
+  \frac{\delta S}{\delta \omega^i_{\ \mu}}\left(\frac{-a}{2(\alpha_3\alpha_4 - a^2)}\,\eta^{ij}\, \epsilon_{\mu\nu\rho}\right) \frac{\delta S}{\delta b^j_{\ \nu}} + \frac{\delta S}{\delta \omega^i_{\ \mu}}\left(\frac{\alpha_4}{2(\alpha_3\alpha_4 - a^2)}\,\eta^{ij}\, \epsilon_{\mu\nu\rho}\right) \frac{\delta S}{\delta b^j_{\ \nu}} &= 0.
\end{align}
The last four terms, proportional to square of Euler derivatives, cancel each other due to antisymmetry of their coefficients. The part surviving is just the missing Poincar\'{e} identity $R_\rho=0$ \eqref{Noether PGT EC 2}. Thus there exist only one set of independent Noether identities.

The antisymmetric terms in the Noether identities \eqref{PGT Gauge Inv MB} again point toward presence of trivial gauge symmetries. To check explicitly, we first write down the general trivial gauge symmetry structure appropriate for the MB model
\begin{align}
\label{trivial gauge MB}
\begin{aligned}
\delta b^i_{\ \mu} &= \Lambda_{\left( b^i_{\ \mu},\, b^j_{\ \nu} \right)} \,\frac{\delta S}{\delta b^j_{\ \nu}}\ +\ \Lambda_{\left( b^i_{\ \mu},\, \omega^j_{\ \nu} \right)} \,\frac{\delta S}{\delta \omega^j_{\ \nu}}\\
\delta \omega^i_{\ \mu} &= \Lambda_{\left( \omega^i_{\ \mu},\, b^j_{\ \nu} \right)} \,\frac{\delta S}{\delta b^j_{\ \nu}} \ +\ \Lambda_{\left( \omega^i_{\ \mu},\, \omega^j_{\ \nu} \right)} \,\frac{\delta S}{\delta \omega^j_{\ \nu}}\\
\end{aligned}
\end{align}
Comparing this with \eqref{symm mapped MB}, we write can down the $\Lambda$ matrix below
\begin{align}
\label{Lambda MB}
\begin{aligned}
\Lambda_{\left( b^i_{\ \mu},\, b^j_{\ \nu} \right)} &= \frac{\alpha_3}{2(\alpha_3\alpha_4 - a^2)}\,\eta^{ij}\,\xi^\rho \epsilon_{\mu\nu\rho} \qquad &
	\Lambda_{\left( b^i_{\ \mu},\,\omega^j_{\ \nu} \right)} &= \frac{-a}{2(\alpha_3\alpha_4 - a^2)}\,\eta^{ij}\,\xi^\rho \epsilon_{\mu\nu\rho} \\
\Lambda_{\left( \omega^i_{\ \mu},\, b^j_{\ \nu} \right)} &= \frac{-a}{2(\alpha_3\alpha_4 - a^2)}\,\eta^{ij}\,\xi^\rho \epsilon_{\mu\nu\rho} \qquad &
	\Lambda_{\left( \omega^i_{\ \mu},\, \omega^j_{\ \nu} \right)} &= \frac{\alpha_4}{2(\alpha_3\alpha_4 - a^2)}\,\eta^{ij}\,\xi^\rho \epsilon_{\mu\nu\rho}
\end{aligned}
\end{align}
The antisymmetry of this structure is easy to verify. We will just demonstrate one component
\begin{align}
\label{antisymm check MB}
\Lambda_{\left( b^i_{\ \mu},\, b^j_{\ \nu} \right)} &= \frac{\alpha_3}{2(\alpha_3\alpha_4 - a^2)}\,\eta^{ij}\,\xi^\rho \epsilon_{\mu\nu\rho}  \nonumber\\
&= - \frac{\alpha_3}{2(\alpha_3\alpha_4 - a^2)}\,\eta^{ji}\,\xi^\rho \epsilon_{\nu\mu\rho} \nonumber \\
&= - \Lambda_{\left( b^j_{\ \nu},\, b^i_{\ \mu} \right)} .
\end{align}
So the two symmetries $\delta_{\scriptscriptstyle G}$ and $\delta_{\scriptscriptstyle PGT}$ differ only by a trivial gauge symmetry which is of no physical importance. The Poincar\'{e} transformations are indeed recovered by the hamiltonian mechanism. An important point to be noted from the analysis of this model is that the hamiltonian symmetries \eqref{symm MB} of this model were different from those of the Einstein-Cartan theory \eqref{symm G EC}. However we could nevertheless recover the Poincar\'{e} symmetries from both of these. The particular difference in details between the models (various terms in the action along with their coupling constants) got manifested only through trivial gauge symmetries.

\section{Discussions}
\label{Sec:Disc}

We have shown in this paper that the Dirac hamiltonian construction indeed reproduces the Poincar\'{e} symmetries in different models of gravity. We have analysed the Einstein-Cartan action and a more generalised form of a Mielke-Baekler type action with a cosmological term, both in 3-dimensions. The Noether identities corresponding to the two sets of symmetries, hamiltonian gauge and Poincar\'{e}, were shown to be the same, modulo antisymmetric cancelling terms proportional to square of Euler derivatives. Using these Noether identities, we derived a map between the two sets of gauge parameters. After using the map, we demonstrated that the difference in the hamiltonian gauge symmetries and the Poincar\'{e} symmetries was just trivial gauge transformations, characterised by coefficients antisymmetric under exchange of fields. We have explicitly found out the coefficient matrices for both Einstein-Cartan and its Mielke-Baekler type generalisation.

Since trivial gauge symmetries are of no physical importance, the Poincar\'{e} symmetries are indeed recovered through the canonical procedure. This feature should persist in all the different diffeomorphism invariant theories of interest and shows the importance of understanding and handling trivial gauge symmetries.

We have shown how the lagrangian and hamiltonian formulations complement each other and how their unified application is of great importance. Analysis of the Noether identities arising in the lagrangian formulation helps us to construct the map between gauge parameters present in the hamiltonian and Poincar\'{e} gauge transformations. This map, at the hamiltonian level, can only be guessed through an (in general case, a rather difficult) exercise of inspection and trial. In the lagrangian procedure, however, the process is much more straightforward and systematic. It is noteworthy that the map is model independent, i.e. it is the same in both examples studied here. This universal nature reveals a unifying feature among the hamiltonian gauge symmetries, a fact that is not otherwise transparent. Indeed, contrary to Poincar\'{e} gauge transformations, the structure of hamiltonian gauge transformations are distinct for distinct models.

Finally, let us recall the role of trivial gauge transformations at the quantum level. This is relevant since gauge symmetries are important in the process of quantisation. The classical gauge symmetries of the action are now replaced by the quantum (Becchi-Rouet-Stora-Tyutin or BRST) symmetries of the quantum effective action $(\Gamma)$. For general gauge theories it was shown \cite{Alexandrov:1998uy} that the set of local symmetries of $\Gamma$ comprise of the quantum gauge transformations, trivial gauge transformations and transformations induced by background fields. Taking a linear combination of all three symmetries, it is possible to find a simple or a standard form. Indeed, adopting this approach the classical gauge transformations for Yang Mills theory were reproduced in \cite{Alexandrov:1998uy}.


\end{document}